\documentclass[12pt]{article}

\usepackage{amsmath}
\usepackage{amssymb}
\usepackage{graphicx, subfig}
\usepackage{url}
\usepackage{multirow}
\usepackage{authblk}
\usepackage{mathtools}
\usepackage{hyperref}
\usepackage[margin=1.25in]{geometry}
\usepackage{natbib}

\usepackage{adjustbox}

\newcommand\independent{\protect\mathpalette{\protect\independenT}{\perp}}
\def\independenT#1#2{\mathrel{\rlap{$#1#2$}\mkern2mu{#1#2}}}

\usepackage{tikz}
\usetikzlibrary{shapes.geometric, arrows}
\tikzstyle{box} = [minimum width=1.0cm, minimum height=1.0cm, text centered, draw=black,fill=white]
\tikzstyle{oval} = [minimum width=1.0cm, minimum height=1.0cm, text centered, draw=black,fill=white]
\tikzstyle{arrow} = [thick,->,>=stealth,line width=1]

\begin{document}

\title{Hierarchical causal variance decomposition for institution and provider comparisons in healthcare}

\author[1]{ Bo~Chen}
%\author[2]{Keith A.~Lawson}
%\author[2]{Antonio~Finelli}
\author[1]{Olli~Saarela\thanks{Correspondence to: Olli Saarela, Dalla Lana School of Public Health, 155 College Street, Toronto, Ontario M5T 3M7, Canada. Email: \texttt{olli.saarela@utoronto.ca}}}

\affil[1]{Dalla Lana School of Public Health, University of Toronto}
%\affil[2]{Princess Margaret Cancer Centre, University Health Network}

\maketitle

\begin{abstract}
Disease-specific quality indicators (QIs) are used to compare institutions and health care providers in terms processes or outcomes relevant to treatment of a particular condition. In the context of surgical cancer treatments, the performance variations can be due to hospital and/or surgeon level differences, creating a hierarchical clustering. We consider how the observed variation in care received at patient level can be decomposed into that causally explained by the hospital performance, surgeon performance within hospital, patient case-mix, and unexplained (residual) variation. For this purpose, we derive a four-way variance decomposition, with particular attention to the causal interpretation of the components. For estimation, we use inputs from a mixed-effect model with nested random hospital/surgeon-specific effects, and a multinomial logistic model for the hospital/surgeon-specific patient populations. We investigate the performance of our methods in a simulation study.

\noindent{\bf Keywords: causal inference, variance decomposition, quality indicator, nested random effects model} 
\end{abstract}

\newpage

\section{Introduction}
\subsection{Background}

Data on health care utilization and patient outcomes have multilevel structure, with clusters formed for example by administrative subregions, referral networks, hospitals, and physicians \citep{daniels1999hierarchical}. Quantifying the between cluster variation in processes of care and outcomes can reveal quality of care related issues, motivating the practice of hospital or provider profiling, with reviews of statistical approaches given for example by \citet{goldstein:1996,Shahian:2008,racz:2010}. Some of the modeling approaches are aimed at identifying outlier clusters \citet{farrell2010outlier}, while others focus on quantifying and explaining the sources of variation \citep{hawley2006correlates}. We will discuss these issues in the context of disease-specific quality indicators (QIs) for surgical care of kidney cancer \citep{wood:2013,lawson2017impact}, where the clustering of interest are surgeons nested within hospitals. The QIs we consider are either process type, capturing variations in care delivered, or outcome type, capturing variation in patient outcomes \citep{donabedian:1988}.

Although it is possible that some surgeons operate in multiple hospitals, a hierarchical clustering can be constructed by considering each hospital-surgeon combination as a separate category.
To introduce some notation, let $Y \in \mathbb{R}$ represent the observed process or outcome experienced by a given patient, and $X = (X_1, \ldots, X_p)$ represent a vector of patient characteristics necessary for case-mix adjustment in the comparisons. Also, let $Z \in \{1,...,m\}$ indicate the hospital in which the patient was actually treated, and $S \in \{1,...,h_z\}$ the surgeon that treated the patient in hospital $z$. The treatment received/outcome of the patient can be modeled through a generalized linear model of the form
\begin{equation}\label{eq:glm}
E[Y \mid Z=z, S=s, X=x]= g^{-1}\left(\alpha_0+\alpha_{z} +\gamma_{zs}+\beta' x \right),
\end{equation}
where $g$ is the link function. The model can be made identifiable by setting the fixed effects $\alpha_1 = 0$ and $\gamma_{z1} = 0$, $z = 1, \ldots m$, or alternatively through random effects by taking them to be IID as $\alpha_{z} \sim N(0, \tau^2)$ and $\gamma_{zs} \sim N(0, \kappa^2)$. The covariance between the hospital and surgeon effects is 0 due to the two-level categories being nested rather than crossed, this model is referred to as a nested random effects model \citep{Ragnar:1986,Nicholas:1987,Sophia:2005}. It also involves the additional assumption that the random effects are independent of the individual-level characteristics \citep{Joseph2014,Paul:2015}. The choice between the fixed effect and random effect formulations depends in part on the numbers of clusters at different levels and numbers of observations per cluster. With large number of clusters, some of these small, the random effect model can provide more stable estimation due to the shrinkage effect for the small clusters, which may be desirable even if the distributional assumption on the random effect is violated. Model \eqref{eq:glm} assumes the absence of interactions between the cluster effects and the individual-level characteristics, but this can be relaxed by allowing for the interactions, which can again be either fixed or random \citep{Andrew:2019}. While in some contexts models such as \eqref{eq:glm} are fitted to estimate the effects $\beta$ of the individual-level characteristics, in the hospital/provider profiling context such a model is typically used to estimate the cluster effects while adjusting for the case-mix factors $X$; predictions from the model can be used to calculate directly standardized estimates of the hospital/provider specific means. The problem of directly standardized comparisons between hospitals of can be framed in a causal inference framework using potential outcomes, as outlined by \citep{varewyck:2014}. 

Another use for models such as \eqref{eq:glm} is to quantify how much variation in the outcome is explained by the cluster-level effects. In the present context this answers the question of whether quality of care differences exist in the health care system overall, in particular, adjusted for patient case-mix, whether similar kinds of patients receive different level of care. Demonstrating such variation is often the first step of validating a proposed process or outcome measure as a QI. In the case of identity link, or at the link function scale, an $X$-conditional variance decomposition can be directly given in terms of the random effect variance parameters \citep{Merlo:2006}. More generally at the outcome scale, variance decompositions can still be calculated making use of model-based predictions. However, generally there exists several alternative variance decompositions depending on the order of conditioning on the variables \citep{bowsher:2012}. In \citet{Chen:2019} we demonstrated that a certain ordering of the variables results in a variance decomposition where the between-hospital component can be given a causal interpretation. We also analyzed hospital-level variation in the quality of surgical care of kidney cancer in Ontario, Canada. The proposed quality indicators we considered were the proportion of partial (versus radical) nephrectomies among stage T1a nephrectomy patients, the same proportion restricted to the subpopulation of patients with chronic kidney disease or its risk factors diabetes or hypertension, minimally invasive surgery among T1-T2 radical nephrectomy patients, and readmission within 30 days of the surgery for T1-T4 radical nephrectomy patients. The first three of these are process type, while readmission is an outcome. While we found significant between-hospital and case-mix variation in several of the indicators, also the residual variances were large, raising the question of how much within-hospital between-surgeon variation is captured by these indicators. This motivates us to further develop a causal variance decomposition and corresponding estimators for hierarchical clusterings, for the purpose of quantifying the contribution of hospital and surgeon level effects while adjusting for patient case-mix. This requires introduction of potential outcomes notation for multiple levels of exposures, which can be adapted from instrumental variable \citep{angrist1996identification} and causal mediation analysis \citep{VanderWeele:2009b, VanderWeele:2014} literature.

\subsection{Objectives}

Based on the objectives motivated above, the structure of the paper is as follows. In Section \ref{Notation and assumptions}, we adapt potential outcomes notation to represent multiple nested exposure levels, and state the necessary assumptions for unconfounded comparisons between the levels. In Section \ref{Four-way decomposition}, we generalize the three-way causal variance decomposition of \citet{Chen:2019} to a four-way decomposition capturing variance components due to patient case-mix, hospitals' performance, surgeons' performance, and unexplained variation, and discuss its causal interpretation. In Section \ref{Hypothetical}, we connect the variance decomposition to measures of intra-class correlation. We propose an estimation method based on nested random-effect models in Section \ref{section:estimation} and study its properties in a simulation study in Section \ref{section:simulation}. We end with a discussion on limitations and future research directions in Section \ref{section:discussion}.

\section{Proposed measures}

\subsection{Notation and assumptions}\label{Notation and assumptions}

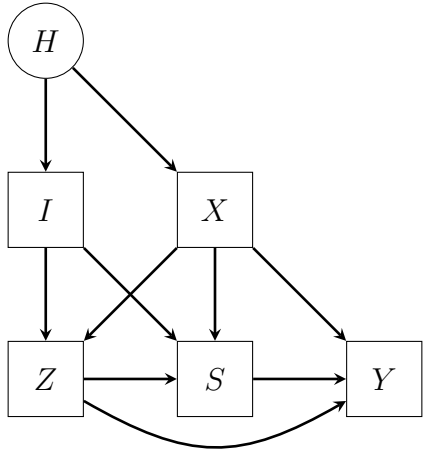
\begin{figure*}[!h]
\begin{center}
\begin{tikzpicture} [node distance=2.25cm]
\node(x) [box] {$X$};
\node(s) [box, below of=x, xshift=0.0cm] {$S$};
\node(z) [box, below of=x, left of=x, xshift=0.0cm] {$Z$};
\node(y) [box, below of=x, right of=x] {$Y$};
\node(v) [box, left of=x] {$I$};
\node(h) [ellipse, oval, left of=x, above of=x, xshift=0.0cm] {$H$};
\draw [arrow] (x) -- (z);
\draw [arrow] (x) -- (y);
\draw[arrow](z)  to [bend right,looseness=1.2] (y);
\draw [arrow] (h) -- (x);
\draw [arrow] (h) -- (v);
\draw [arrow] (v) -- (z);
\draw [arrow] (z) -- (s);
\draw [arrow] (s) -- (y);
\draw [arrow] (x) -- (s);
\draw [arrow] (v) -- (s);
\end{tikzpicture}
\end{center}
\caption{Causal mechanism for hospital ($Z$) assignment,  surgeon ($S$) assignment and a process of care ($Y$). $X$ represents a vector of potential confounders relevant to the case-mix adjustment, while $I$ represents instrumental variables that predict the hospital assignment but are not confounders. $H$ represents latent history that can influence $I$ and $X$ but is not in itself confounder.}\label{figure:dag}
\end{figure*}

We suppress the individual level index $i$, and let $Y \in \mathbb{R}$ represent the observed process or outcome experienced by a given patient, used to construct a QI. Let $Y(z, s) \in \mathbb{R}$ represent the counterfactual outcomes of the same patient received care via surgeon $s \in \{1 , \ldots, h_z\}$ operating in a hospital $z \in \{1, \ldots, m\}$. Let further $X=(X_{1}, \ldots, X_{p})$ be a vector of covariates relevant to the case-mix adjustment, which include demographic, comorbidity, and disease progression information. Let $Z \in \{1, \ldots, m\}$ indicate the hospital where the patient was actually treated, and $S \in \{1 , \ldots, h_z\}$ the surgeon that operated the patient. Let further $S(z) \in\{1,..,h_z\}$ indicate the surgeon that potentially operates the patient, if referred to hospital $z$. The observed variables $Y$ and $S$ are linked to their potential counterparts under the counterfactual consistency/stable unit treatment value assumption (SUTVA), by $Y= Y(Z,S(Z))$ and $S = S(Z)$. Causal inferences on the hospital and surgeon effects are possible under the assumption of strong ignorability of the joint hospital and surgeon assignment mechanism, which states that $0 < P(Z = z, S = z \mid X = x) < 1$ for all $z \in \{1, \ldots, m\}$, $s \in \{1 , \ldots, h_z\}$ and $x$ (positivity) and $Y(z, s) \independent  (Z,S) \mid X$ (conditional exchangeability) \cite{rosenbaum:1983, Miguel:2006}. We note that positivity is assumed only over the observed hospital/surgeon combinations, to make the levels nested rather than crossed. We  introduce the shorthand notations $g(s;z,X) \equiv P(S=s \mid Z=z, X)$ and $e(z; X) = P(Z = z \mid X)$ for the corresponding assignment probabilities. The hypothesized causal relationships are illustrated in the directed acyclic graph (DAG) in Figure \ref{figure:dag}. Here $I$ and $H$ represent additional factors that influence the hospital/surgeon assignment and the patient characteristics, but are not confounders. In particular, adjustment for the instrumental variables $I$ would likely lead to positivity violations.

We note that since $Y(z, S(z)) = Y(z)$, the above notation reduces to the hospital-level potential outcome notation used by for example \citep{varewyck:2014} and \citet{Chen:2019}. For this one-level clustering, in \citet{Chen:2019} we derived a variance decomposition 
\begin{align}\label{intro1}
V[Y] &= V_{X}\left\{\sum_{z}E(Y(z) \mid X) e(z; X)\right\} \nonumber \\
&\quad + E_{X} \left \{ \sum_{z}\left[E(Y(z) \mid X)  - \sum_{z'} E(Y(z') \mid X) e(z'; X) \right]^2 e(z; X) \right \} \nonumber \\
&\quad + E_{X} \left \{ \sum_{z} V(Y(z) \mid X) e(z; X) \right \}\\
&= \textrm{variance explained by the patient case-mix} \nonumber \\
&\quad + \textrm{average variance causally explained by the between-hospital differences in performance} \nonumber \\
&\quad\quad \textrm{conditional on case-mix} \nonumber \\
&\quad + \textrm{residual variance}.\nonumber 
\end{align}
Here the second variance component captures the average squared differences from the average level of care for similar patients between the hospitals. However, it does not capture between provider variation within the hospitals, which is included in the residual variance. In the following we derive a four-way decomposition that introduces a new term to capture the within hospital between provider variation.

\subsection{Four-way decomposition for observed varation in care received}\label{Four-way decomposition}

Under counterfactual consistency/SUTVA, we have $V[Y]=V[Y(Z,S(Z))]$. We begin with the two-way variance decomposition
\begin{equation} \label{eq1}
V[Y(Z,S(Z))]=V_{X}[E(Y(Z,S(Z)) \mid X)]+E_{X}[V(Y(Z,S(Z))) \mid X)].
\end{equation}
In the Eqution (\ref{eq1}), the first term can further write
\begin{align}\label{eq2}
V_{X}[E(Y(Z,S(Z)) \mid X)]=&V_{X}[ E_{Z \mid X}[E(Y(Z,S(Z)) \mid Z,X)]]\nonumber \\
=&V_{X}\{ E_{Z \mid X}[E_{S(Z) \mid Z,X}[E(Y(Z,S(Z)) \mid S(Z), Z,X)]]\}
\end{align}
In the latter term, we can further write
\begin{align}\label{eq3}
\MoveEqLeft E_{X}[V(Y(Z,S(Z)) \mid X)] \nonumber \\
&= E_{X}[V_{Z \mid X}[E(Y(Z,S(Z)) \mid Z, X)]] \nonumber \\
&\quad+ E_{X}[E_{Z \mid X}[V(Y(Z,S(Z)) \mid Z, X)]] \nonumber \\
&= E_{X}\{V_{Z \mid X}[E_{S(Z) \mid Z, X}[E(Y(Z,S(Z)) \mid S(Z), Z, X)]]\} \nonumber \\
&\quad+E_{X}\{E_{Z \mid X}[V_{S(Z) \mid Z, X}[E(Y(Z,S(Z)) \mid S(Z), Z, X)]]\}\nonumber \\
&\quad+E_{X}\{E_{Z \mid X}[E_{S(Z) \mid Z, X}[V(Y(Z,S(Z)) \mid S(Z), Z, X)]]\}.
\end{align}
Substituting these Equations (\ref{eq2}) and (\ref{eq3}) into Equation (\ref{eq1}), we obtain 
\begin{align}\label{eq4}
V[Y(Z,S(Z))] &= V_{X}\{ E_{Z \mid X}[E_{S(Z) \mid Z,X}[E(Y(Z,S(Z)) \mid S(Z), Z,X)]]\}\nonumber \\
&\quad+E_{X}\{V_{Z \mid X}[E_{S(Z) \mid Z, X}[E(Y(Z,S(Z)) \mid S(Z), Z, X)]]\}\nonumber \\
&\quad+E_{X}\{E_{Z \mid X}[V_{S(Z) \mid Z, X}[E(Y(Z,S(Z)) \mid S(Z), Z, X)]]\}\nonumber \\
&\quad+E_{X}\{E_{Z \mid X}[E_{S(Z) \mid Z, X}[V(Y(Z,S(Z)) \mid S(Z), Z, X)]]\}.
\end{align}
Due to strong ignorability assumption, we have 
\begin{equation*}
E(Y(z, s) \mid S=s, Z=z,X) = E(Y(z, s) \mid X) 
\end{equation*}
and
\begin{equation*}
V(Y(z, s) \mid S=s, Z=z,X) = V(Y(z, s) \mid X).
\end{equation*}
Hence, we can obtain the following result in terms of the potential outcomes:
\begin{align}\label{eq7}
V[Y] &= V_X \bigg \{\sum_{z} \sum_s E(Y(z, s) \mid X)g(s;z,X) e(z; X)\bigg \}\nonumber\\
&\quad+E_{X}\bigg\{ \sum_{z} \bigg[ \sum_s E(Y(z, s) \mid X)g(s;z,X) \nonumber\\
&\qquad\qquad\qquad\quad-\sum_{z'} \sum_s E(Y(z', s) \mid X)g(s;z',X)e(z'; X) \bigg]^2 e(z; X)\bigg\}\nonumber\\
&\quad+E_{X}\bigg\{\sum_{z} \bigg[ \sum_s \bigg(E(Y(z, s) \mid X)-\sum_{s'}E(Y(z, s') \mid X)g(s';z,X) \bigg)^2 g(s;z,X) \bigg] e(z; X) \bigg\}\nonumber\\
&\quad +E_{X}\bigg \{\sum_{z}\sum_s V(Y(z, s) \mid X)g(s;z,X)e(z; X)\bigg \}.
\end{align}
The interpretation is
\begin{equation*}
\begin{split}
V[Y]& = \textrm{total observed variance in care received}\\
&= \textrm{variance explained by the patient case-mix} \\
&\quad + \textrm{average variance causally explained by the between hospital} \\
&\quad \quad \textrm{differences in performance conditional on case-mix} \\
&\quad +\textrm{average variance causally explained by the surgeon performance}\\
&\quad \quad \textrm{conditioning on patient case-mix and hospital performance} \\
&\quad + \textrm{unexplained (residual) variance}.
\end{split}
\end{equation*}
We note that the first and second term in \eqref{eq7} are equivalent to the first and second term in \eqref{intro1}. This also implies that the third, residual, variance component in \eqref{intro1} is equivalent to the sum of the third and fourth components in \eqref{eq7}, meaning that the additional term in \eqref{eq7} is a result of splitting the residual variance in \eqref{intro1}. We will consider the second and the third terms as the causal quantities of interest. These have a causal interpretation, as further discussed in Section \ref{interpretation}, and can be linked to the observable quantities under the causal assumptions, working backwards from \eqref{eq7}. In the second term, the hospital performance is compared to the average level across all the hospitals for a patient with characteristics $X$ and then averaged over the patient population. In the third term, the surgeon performance is compared to the average level of the surgeons in the same hospital for a patient with characteristics $X$, and then averaged over the hospitals and the patient population. 

There are two possible approaches to estimate the variance components in \eqref{eq7}. In the first approach, directly based on the factorization in $\eqref{eq7}$, we can estimate them based on modeling $E[Y\mid S,Z, X]$, $P(S \mid Z,X)$, $P(Z\mid X)$ and using the empirical distribution of $X$. Because the four-way variance decomposition can be also expressed as
\begin{align}\label{eq8}
V[Y] &= V_{X} [E(Y\mid X)]\nonumber\\
&\quad+ E_{Z,X}\big\{[E(Y\mid Z,X)-E(Y\mid X) ]^2\big\}\nonumber\\
&\quad+E_{S,Z,X}\big\{[E(Y\mid S,Z,X)-E(Y\mid Z,X)]^2\big\}\nonumber\\
&\quad+E_{S,Z,X}[V(Y\mid S,Z,X)],
\end{align}
the variance components can be also estimated based on modeling $E[Y\mid S(Z),Z, X]$, $E[Y\mid Z,X]$, $E[Y\mid X]$ and using the empirical distribution of $(S, Z, X)$. We will discuss both approached in Section  \ref{section:estimation}; the former is based on factorization of the likelihood and can be used to construct an approximate Bayesian inference procedure.

\subsection{Causal interpretation of the decomposition}\label{interpretation}

To better understand the causal interpretation of the variance decomposition, we considered a special case with two hospitals with two surgeons each. The interpretation of the case-mix and between-hospital components is unchanged and was already discussed by \citet{Chen:2019}, so we focus on the interpretation of the within-hospital between-surgeon component. With two hospitals with indexed by $z=1, 2$, and two surgeons in each indexed by $s=1,2$, we denote the hospital assignment probability with $P(Z=1\mid X)=e(1; X)=e(X)$, with $P(Z=2\mid X) = 1-e(X)$, and the surgeon assignment probability within hospital with $P(S=1\mid Z=z, X)=g(1;z,X)=g(z,X)$, with $P(S=2\mid Z=z, X) = 1-g(z,X)$. Now the third term \eqref{eq7} becomes
\begin{align}\label{eq9}
\MoveEqLeft E_X\Big\{e(X) g(1,X)(1-g(1,X))\left[E(Y(1,1)\mid X)-E(Y(1,2)\mid X)\right]^2\nonumber\\
&+ (1-e(X)) g(2,X)(1-g(2,X))\big[E(Y(2,1)\mid X)-E(Y(2,2)\mid X)\big]^2 \Big\}\nonumber\\
&=E_X\Big\{ e(X) V(S\mid Z=1,X)\left[E(Y(1,1)\mid X)-E(Y(1,2)\mid X)\right]^2\nonumber\\
&\quad\qquad+ (1-e(X))V(S\mid Z=2,X)\left[E(Y(2,1)\mid X)-E(Y(2,2)\mid X)\right]^2 \Big\}.
\end{align}
We consider three scenarios to illustrate the causal interpretation of \eqref{eq9} via the relationship of $X, Z, S$ and $Y$ in Figure \ref{figure:dag}.
\begin{description}
\item[Scenario 1.] In the absence of the arrow $X \rightarrow Y$, which implies $Y \independent X \mid (Z,S)$ and $E(Y(z,s)\mid X)=E(Y(z,s))$, \eqref{eq9} becomes 
\begin{align*}
&\left[E(Y(1,1))-E(Y(1,2))\right]^2 E_X\left[ e(X) V(S\mid Z=1,X)\right]\\
&+ \left[E(Y(2,1))-E(Y(2,2))\right]^2 E_X\left[(1-e(X))V(S\mid Z=2,X) \right].
\end{align*}
The first multiplicative terms represent squared pairwise causal contrasts. This is multiplied by the second terms, the magnitude of which depends on the volume of patients of type $X$ in each hospital, as well as the variation in the surgeon assignment for patients of type $X$. The latter is maximized when both surgeons treat similar patient populations. If the surgeons specialize on treatment of specific kinds of patients, so that there is no overlap in the patient populations treated by the two surgeons, the positivity assumption is violated and the corresponding component is equal to 0. Thus, for between-surgeon performance differences to manifest through this variance component, there must be some overlap in the patient population they treat.

\item[Scenario 2.] In the absence of the arrow $X \rightarrow S$, $X \rightarrow Z$, $I \rightarrow S$, $I \rightarrow Z$, which implies $(Z,S) \independent X$, we have $e(X)=e$, $g(1,X)=g(1)$, and $g(2,X)=g(2)$. Now \eqref{eq9} becomes
\begin{align*}
& e V(S\mid Z=1) E_X \bigg\{\big[ E(Y(1,1)\mid X)-E(Y(1,2)\mid X)\big]^2 \bigg\}\\
&+(1-e) V(S\mid Z=2) E_X \bigg\{\big[ E(Y(2,1)\mid X)-E(Y(2,2)\mid X)\big]^2 \bigg\}.
\end{align*}
We note that under this completely randomized setting the magnitude of the causal effects are proportional to the terms
\begin{align*}
\MoveEqLeft E_X \left\{\left[ E(Y(z,1)\mid X)-E(Y(z,2)\mid X)\right]^2 \right\} \\
&= E_X \left\{E(Y(z,1) - Y(z,2)\mid X)^2 \right\} \\
&= V_X\left[E(Y(z,1) - Y(z,2)\mid X)\right] + E_X \left\{E(Y(z,1)-Y(z,2)\mid X) \right\}^2 \\
&= V_X\left[E(Y(z,1) - Y(z,2)\mid X)\right] + E[Y(z,1)-Y(z,2)]^2,
\end{align*}
that is, proportional to the sum of the variance of the covariate conditional causal effects and the squared population average causal effect. The former captures effect modification by the patient characteristics, showing that any effect modification adds to the measure, rather than canceling out, and the latter captures the overall performance difference. The resulting variance component, as expressed for two levels being compared, has similarities to the causal interpretation recently given to the model reliance metric used to measure variable importance in machine learning contexts \citep{fisher2019all}. Our results show that a similar kind of effect measure can be derived through a variance decomposition argument under a randomized assignment, and generalize this from two to multiple exposure levels being compared.  Since the variance component under a randomized assignment mechanism may be of interest in itself as a causal quantity, in Section \ref{Hypothetical} we show that it can always be derived and estimated under a hypothetical target assignment mechanism of interest regardless of the actual mechanism that assigns patients for hospitals and surgeons.
%\item[Scenario 3.] In the absence of the arrow $Z \rightarrow Y$, which implies $Y \independent Z \mid (S, X)$.
\item[Scenario 3.] In the absence of the arrow $S \rightarrow Y$, which implies $Y \independent S \mid (Z, X)$, we have $E(Y(1,1)\mid X)=E(Y(1,2)\mid X)$ and $E(Y(2,1)\mid X)=E(Y(2,2)\mid X)$, and the between-surgeon component is zero, as it should it the absence of individual-level causal effects.
\end{description}

\subsection{Hypothetical assignment mechanism}\label{Hypothetical}

The variance decomposition \eqref{eq7} was derived for the observed marginal variance of the outcome, which is why it depends on the mechanism that assigns patients to hospitals and surgeons, including the hospital and surgeon volume. Alternatively, we can derive a variance decomposition under a hypothetical ``randomized'' assignment, where for example each hospital/surgeon treats similar kind of patient population, and/or similar patient volume. This will also allow us to derive a connection between the causal variance decomposition and well-known intra-class correlation measures. Let $A \in \{1, \ldots, m\}$ and $B(A) \in s \in \{1 , \ldots, h_a\}$ be hospital and surgeon assignments randomly drawn with specified probabilities 
$\tilde e(a; X) = P(A = a \mid X)$ and $\tilde g(b;a,X) \equiv P(B=b \mid A=a, X)$, chosen such that $(A, B)\independent (Z, S) \mid X$, $0< P(A=a, B=b\mid X=x)<1$ and $Y(a,b) \independent (A,B)\mid X$. Here choosing for example $\tilde e(a; X) = P(Z = a)$ and $\tilde g(b;a,X) = P(S=b \mid Z=a)$ would correspond to a mechanism where each hospital/surgeon treats similar patient population, but retaining the original patient volumes. Choosing $\tilde e(a;X) \equiv 1/m$ and $\tilde g(b;a,X) = 1/h_a$ would also mean setting the volumes to be the same.

For the variance under the hypothetical assignment mechanism we get the decomposition
\begin{align}\label{hypothetical}
\MoveEqLeft V[Y(A,B(A))] \nonumber \\
&= V_X \bigg \{\sum_{a} \sum_b E(Y(a, b) \mid X)\tilde g(b;z,X) \tilde e(a; X)\bigg \}\nonumber\\
&\quad+E_{X}\bigg\{ \sum_{a} \bigg[ \sum_b E(Y(a, b) \mid X)\tilde g(b;a,X) \nonumber\\
&\qquad\qquad\qquad\quad-\sum_{a'} \sum_b E(Y(a', b) \mid X) \tilde g(b;a',X) \tilde e(a'; X) \bigg]^2 \tilde e(a; X)\bigg\}\nonumber\\
&\quad+E_{X}\bigg\{\sum_{a} \bigg[ \sum_b \bigg(E(Y(a, b) \mid X)-\sum_{b'}E(Y(a, b') \mid X)\tilde g(b';a,X) \bigg)^2 \tilde g(b;a,X) \bigg] \tilde e(a; X) \bigg\}\nonumber\\
&\quad +E_{X}\bigg \{\sum_{a}\sum_b V(Y(a, b) \mid X)\tilde g(b;a,X)\tilde e(a; X)\bigg \}.
\end{align}
Because under the above assumptions, we have
\begin{align*}
E[Y(a, b) \mid A=a, B=b, X]
&=  E[Y(a, b) \mid X] \\
&= E[Y(a,b) \mid Z=a,S=b, X] \\
&= E[Y \mid Z=a,S=b, X],
\end{align*}
and because $\tilde e(a; X)$ and $\tilde g(b;a,X)$ are fixed quantities, this is also estimable from the observed data on $(Y,Z,S,X)$. 

Under the special case of $\tilde e(a;X) \equiv 1/m$ and $\tilde g(b;a,X) = 1/h_a$ and the linear mixed-effect model
\begin{equation*}
E[Y(a, b) \mid Z=a, S=b, X]=E[Y(a, b) \mid X]=\alpha_0+\alpha_a+\gamma_{ab}+\beta'X,
\end{equation*}
where the nested hospital and surgeon random effects are IID as $\alpha_{a} \sim N(0, \tau^2)$ and $\gamma_{ab} \sim N(0, \kappa^2)$ and residuals distributed as $Y - E[Y \mid Z, S, X] \sim N(0,\sigma^2)$, the second term in \eqref{hypothetical} can be written as
\begin{equation*}
\frac{1}{m}\sum_{a}\left\{\left(\alpha_a-\frac{1}{m}\sum_{a'}\alpha_{a'}\right)+\left(\frac{1}{h_a}\sum_{b}\gamma_{ab}-\frac{1}{m}\sum_{a'}\frac{1}{h_{a'}}\sum_{b}\gamma_{a'b}\right)\right\}^2.
\end{equation*}
Keeping $m$ is fixed and letting $h_a\rightarrow\infty$ for all $a \in \{1, \ldots, m\}$, the terms $\frac{1}{h_a}\sum_{b}\gamma_{ab}$ converge to $E(\gamma_{ab})=0$, and the second term in becomes \eqref{hypothetical}
\begin{equation}\label{hypothetical_term2}
\frac{1}{m}\sum_{a}\bigg\{\alpha_a-\frac{1}{m}\sum_{a'}\alpha_{a'}\bigg\}^2,
\end{equation}
which in turn converges to $V(\alpha_a)=\tau^2$, the variance of the between-hospital effects. Hence, for this variance component we obtain the same result as for the three-way causal variance decomposition proposed by \citet{Chen:2019}. 

The third term in \eqref{hypothetical} can be written as
\begin{equation}\label{hypothetical_term3}
\frac{1}{m}\sum_{a}\frac{1}{h_a}\sum_{b}\left(\alpha_{ab}-\frac{1}{h_a}\sum_{b'}\gamma_{ab'}\right)^2.
\end{equation}
Keeping $m$ fixed and letting $h_a\rightarrow\infty$ for all $a \in \{1, \ldots, m\}$, \eqref{hypothetical_term3} converges to $V(\alpha_{ab})=\kappa^2$, the variance of the within-hospital between-surgeon effects. Therefore, under this special case, the $X$-conditional causal variance decomposition is
\begin{align*}
V(Y(A,B(A)) \mid X) &=  V_{A \mid X}[E_{B(A) \mid A, X}[E(Y(A,B(A)) \mid B(A), A, X)]]\\
&\quad+ E_{A \mid X}[V_{B(A) \mid A, X}[E(Y(A,B(A)) \mid B(A), A, X)]]\\
&\quad+ E_{A \mid X}[E_{B(A) \mid A, X}[V(Y(A,B(A)) \mid B(A), A, X)]]\\
&\rightarrow  \tau^2 + \kappa^2+ \sigma^2 \;\textrm{when} \; h_a\rightarrow\infty \; \forall a\in\{1,..,m\} \; \textrm{and} \; m \rightarrow \infty,
\end{align*}
that is, asymptotically equivalent to the variance decomposition obtained through the model-based random effect and residual variances. Here $(\tau^2+\kappa^2)/(\tau^2 + \kappa^2+ \sigma^2)$ would correspond to the within-hospital within-surgeon intra-class correlation coefficient.

\section{Estimators}\label{section:estimation}

\subsection{Point estimation}\label{Point estimation}

We outline estimators based on the decomposition \eqref{eq7}, which requires fitting hospital, surgeon and case-mix conditional outcome model and case-mix conditional assignment model. The same estimation approach works also for decomposition \eqref{eq9}, but substituting fixed target assignment probabilities in place of the observed ones. 

To model the outcomes, we use a generalized linear mixed model
\begin{equation}\label{mixedmodel}
E[Y(z, s) \mid X;\theta]=E[Y \mid Z=z,S=s, X;\theta]= g^{-1}\left(\alpha_0+\alpha_{z} +\gamma_{zs}+\beta^{'}X \right).
\end{equation}
where  $\theta = (\alpha_0,\alpha_z, \alpha_s, \beta)$ and where the nested hospital and surgeon random effects are taken to be IID as $\alpha_{z} \sim N(0, \tau^2)$ and $\gamma_{zs} \sim N(0, \kappa^2)$. For the joint hospital/surgeon assignment mechanism, we fit a multinomial logistic regression model
\begin{align}
\MoveEqLeft P(Z=z, S=s \mid X;\eta)=  \nonumber \\ 
&\begin{cases}
\frac{1}{1+\sum_{a=2}^{m}\exp(\psi_{a1}+\phi_{a1}^{'}X))+\sum_{b=2}^{h_1}\exp(\psi_{1b}+\phi_{1b}^{'}X))+\sum_{a=2}^m\sum_{b=2}^{h_z} \exp(\psi_{ab}+\phi_{ab}^{'}X)} & z=1, s=1\\
\frac{\exp(\psi_{z1}+\phi_{z1}^{'} X)}{1+\sum_{a=2}^{m}\exp(\psi_{a1}+\phi_{a1}^{'}X))+\sum_{b=2}^{h_1}\exp(\psi_{1b}+\phi_{1b}^{'}X)+\sum_{a=2}^m\sum_{b=2}^{h_z} \exp(\psi_{ab}+\phi_{ab}^{'}X)}& z\neq1,s=1\\
\frac{\exp(\psi_{1s}+\phi_{1s}^{'} X)}{1+\sum_{a=2}^{m}\exp(\psi_{a1}+\phi_{a1}^{'}X))+\sum_{b=2}^{h_1}\exp(\psi_{1b}+\phi_{1b}^{'}X)+\sum_{a=2}^m\sum_{b=2}^{h_z} \exp(\psi_{ab}+\phi_{ab}^{'}X)}& z=1,s\neq1\\
\frac{ \exp(\psi_{zs}+\phi_{zs}^{'}X_i)}{1+\sum_{a=2}^{m}\exp(\psi_{a1}+\phi_{a1}^{'}X))+\sum_{b=2}^{h_1}\exp(\psi_{1b}+\phi_{1b}^{'}X)+\sum_{a=2}^m\sum_{b=2}^{h_z} \exp(\psi_{ab}+\phi_{ab}^{'}X)},& z\neq1,s\neq1
\end{cases}
\label{eq:multi}
\end{align}
where $\eta = \{(\psi_{ab}, \phi_{ab}) : (a, b) \ne (1,1)\}$. Hence, we can obtain
\begin{equation*}
e(z; X, \eta) = P(Z=z \mid X;\eta) = \sum_{s=1}^{h_z} P(Z=z, S=s \mid X;\eta)
\end{equation*}
and
\begin{equation*}
g(s;z,X,\eta) = P(S=s \mid Z=z, X;\eta)  = \frac{P(Z=z, S=s \mid X;\eta)}{P(Z=z \mid X;\eta)}.
\end{equation*}
Alternatively, a multinomial logistic assignment model can be first fitted at hospital level to estimate the quantities $e(z; X, \eta)$, after which surgeon level multinomial assignment models are fitted conditionally on each hospital to estimate $g(s;z,X,\eta)$.

We denote the fitted values for the expected outcomes $\mu_i(z,s;\theta)=E(Y_i \mid Z_i=z, S_i=s, x_i;\theta)$. Under the mixed-effects model these are obtained by using empirical Bayes
prediction for the random hospital and surgeon-level intercepts \citep[e.g][Chapter 7]{skrondal2004generalized}. Further, we denote $V[Y;\theta, \eta] = \omega_1(\theta,\eta) + \omega_2(\theta,\eta) + \omega_3(\theta,\eta) + \omega_4(\theta,\eta)$ for the four terms in the parametrized version of the variance decomposition \eqref{eq7}. The first (case-mix) component can now be estimated by
\begin{align*}
\omega_1(\hat\theta,\hat\eta) = \frac{1}{n-1} \sum_{i=1}^n \bigg\{&\sum_{z}\sum_s \mu_i(z,s;\hat{\theta})g(s;z,x_i,\hat\eta)e(z; x_i, \hat \eta) \\
&- \frac{1}{n} \sum_{i'=1}^n \sum_{z} \sum_s \mu_i(z,s;\hat{\theta})g(s;z,x_{i'},\eta)e(z; x_{i'}, \hat\eta) \bigg\}^2.
\end{align*}
The second (between-hospital) component can be estimated by
\begin{align*}
\omega_2(\hat\theta,\hat\eta) = \frac{1}{n} \sum_{i=1}^n \bigg\{&\sum_{z}\bigg[\sum_s \mu_i(z,s;\hat{\theta})g(s;z,x_i,\hat\eta)\bigg]^2 e(z; x_i, \hat \eta)\\
& -\bigg[\sum_{z}\sum_s \mu_i(z,s;\hat{\theta})g(s;z,x_i,\hat\eta) e(z; x_i, \hat \eta)\bigg]^2\bigg\}.
\end{align*}
The second (between-surgeon) component can be estimated by
\begin{align*}
\omega_3(\hat\theta,\hat\eta) = \frac{1}{n} \sum_{i=1}^n \bigg\{\sum_{z}\bigg[&\sum_s \mu_i(z,s;\hat{\theta})^2 g(s;z,x_i,\hat\eta)\\
&\quad -\bigg(\sum_s \mu_i(z,s;\hat{\theta})g(s;z,x_i,\hat\eta)\bigg)^2\bigg] e(z; x_i, \hat \eta)\bigg\}.
\end{align*}
The fourth (residual) variance can be estimated by subtracting the sum of the above three components from the empirical marginal variance, or alternatively, based on the distributional assumption in the outcome model. In particular, for a binary outcome we have $V(Y_i \mid Z_i=z, S_i=s, x_i;\theta)=\mu_i(z,s;\theta)[1-\mu_i(z,s;\theta)]$, and the residual variance component is given by
\begin{align*}
\MoveEqLeft  \omega_4(\hat\theta,\hat\eta) = \frac{1}{n} \sum_{i=1}^n \bigg\{\sum_{z}\sum_{s}\mu_i(z,s;\hat{\theta})[1-\mu_i(z,s;\hat{\theta})]g(s;z,x_i,\hat\eta) e(z; x_i, \hat \eta)\bigg\}.
\end{align*}
If the parameters can be estimated consistently such that $\hat\theta \stackrel{p}{\rightarrow} \theta$ and $\hat\eta \stackrel{p}{\rightarrow} \eta$, the component estimators $\omega_j(\hat\theta,\hat\eta)$, $j \in \{1,2,3,4\}$, are also consistent by the continuous mapping theorem and applying the law of large numbers for the sample averages over the empirical covariate distribution. We will briefly investigate their asymptotic normality through simulation in Section \ref{section:simulation}, but the variance estimation approach we propose in \ref{Variance_estimation} does not make use of asymptotic normality.
 
Alternatively to the above model-based estimators, as noted in Section \ref{Four-way decomposition}, a semi-parametric estimation procedure is suggested by decomposition \eqref{eq8}, based on three different outcome models. The semi-parametric approach is applicable to the decomposition of the empirical marginal variance, while the model-based approach can also be used to estimate decomposition under hypothetical assignment mechanisms. We will briefly compare the two approaches in the simulation study of Section \ref{section:simulation}. 

\subsection{Variance estimation}\label{Variance_estimation}

The point estimators suggested in Section \ref{Point estimation}, are entirely model-based, conditional on the empirical covariate distribution. Because the model components correspond to the factorization of the likelihood, we can evaluate the uncertainty in the variance component estimates via approximate Bayesian inference. This is based on drawing samples from the joint posterior distribution of the parameters $\theta$ and $\eta$, given by
\begin{align*}
f(\theta, \eta \mid  \mathbf{Y}, \mathbf{Z}, \mathbf{S}, \mathbf{X}) &= \frac{f(\mathbf{Y}, \mathbf{Z}, \mathbf{S} \mid \mathbf{X}, \theta, \eta) f(\theta \mid \mathbf{X}) f(\eta \mid \mathbf{X})}{f(\mathbf{Y}, \mathbf{Z}, \mathbf{S} \mid \mathbf{X})} \\
&=  \frac{f(\mathbf{Y} \mid \theta, \mathbf{Z}, \mathbf{S}, \mathbf{X}) f(\theta \mid \mathbf{X})}{f(\mathbf{Y} \mid \mathbf{Z}, \mathbf{S},\mathbf{X})} 
\frac{f(\mathbf{Z}, \mathbf{S}\mid \eta, \mathbf{X}) f(\eta \mid \mathbf{X})}{f(\mathbf{Z},\mathbf{S} \mid \mathbf{X})} \\
&= f(\theta \mid \mathbf{Y}, \mathbf{Z},\mathbf{S}, \mathbf{X}) f(\eta \mid \mathbf{Z},\mathbf{S}, \mathbf{X}).
\end{align*}
Posterior samples for the variance components can be obtained by sampling $\theta$ and $\eta$ from their posterior distributions, and recalculating $\omega_1(\theta,\eta), \omega_2(\theta,\eta), \omega_3(\theta,\eta)$ and $\omega_4(\theta,\eta)$ for each draw. For the outcome model parameters $\theta$, we approximated the posterior using the parametric bootstrap, by resampling outcomes from the fitted model, refitting the model and calculating new fitted values $\mu_i(z,s;\theta)$. For the assignment model parameters, we used the normal approximation to sample the $\eta$ from a multivariate normal distribution $MVN(\hat{\eta},V(\hat{\eta}))$, where $\hat{\eta}$ is the maximal likelihood estimator and $V(\hat{\eta})$ is the asymptotic variance-covariance matrix via the fitted multinomial logistic regression.

\section{Simulation study}\label{section:simulation}

\subsection{Generating mechanism}

We used simulation to study the properties of the methods proposed in Section \ref{section:estimation}. The objectives for the simulation study were to (a) study the asymptotic properties (consistency, asymptotic normality) of the proposed point estimators, (b) to demonstrate that the new four-way decomposition is consistent with our previously proposed three-way decomposition, and (c) to compare the model-based estimators to the alternative semi-parametric decomposition to verify that both are estimating the same quantity. We used a data-generating mechanism similar to Figure \ref{figure:dag}, omitting the variables $I$ and $H$ for simplicity. The asymptotic behavior of the estimators was studies by varying the total number of hospitals $m$, the total number of surgeons $q$, and the total number of patients $n$. We generated two patient case-mix factors, $X_1\sim N(0,1)$ and $X_2\sim \textrm{Bernoulli}(0.5)$. The hospital ($Z$) and surgeon ($S$) assignments were generated based on multinomial logistic model \eqref{eq:multi}, where the intercepts were generated from $N(0,0.25)$ and coefficients from $N(0,0.5)$. Outcomes were generated from the mean structure
\begin{equation*}
E[Y(z,s) \mid X] = \alpha_{z} +\gamma_{zs}+ X_{1} + 2 X_{2}
\end{equation*}
where the hospital's effects $\alpha_{s}$ and surgeon effects $ \gamma_{zs}$ were generated independently from $N(0,2)$. The continuous outcomes were generated by taking $Y(z,s) = E[Y(z,s) \mid X] + \varepsilon$, where $\varepsilon \sim \textrm{Logistic}(0,1)$, and the binary outcomes by dichotomizing these as $\mathbf 1_{\{Y(z,s) \ge 0\}}$. The observed outcomes were takes to be $Y = Y(Z,S)$. For estimation, we fitted mixed effect logistic models with nested random effects as in \eqref{mixedmodel} and multinomial assignment models as in \eqref{eq:multi}. The resulting estimates for the variance components were compared to the true values calculated under the above specified parameter values.

\subsection{Results}

\begin{figure}[!ht]
\centering
\includegraphics[width=\textwidth]{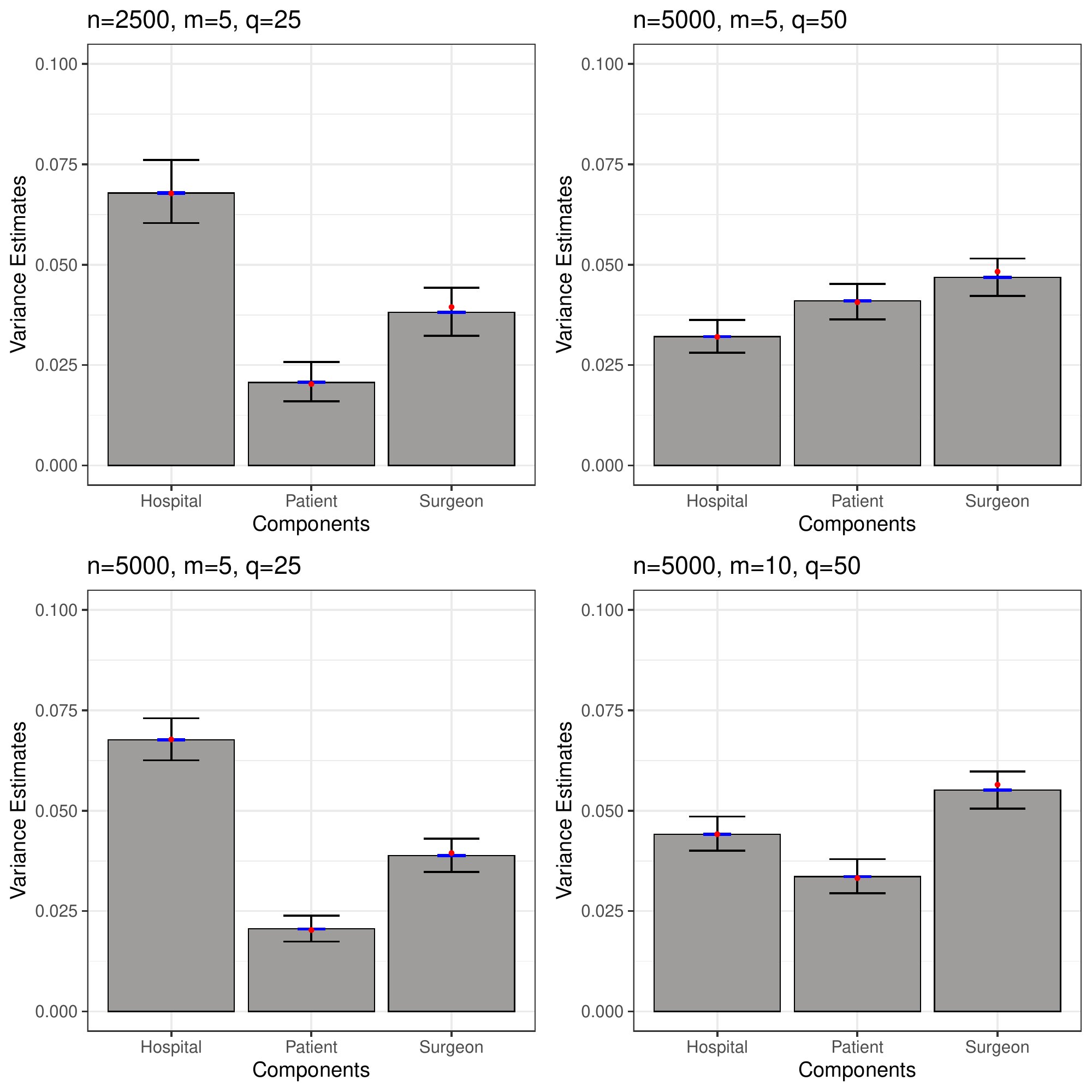}
\caption{Simulated sampling distribution means for the three variance components (without residual variance) under the random-effect model for the binary outcomes under different combinations of $n$ (total number of patients), $m$ (total number of hospitals), and $q$ (total number of surgeons), based on 1000 replications. The red dots indicate the true variances. The $95\%$ quantile interval of the sampling distribution is represented by the black error bar. The $95\%$ confidence interval for the mean is represented by the blue error bar, reflecting the Monte Carlo error in the estimated mean of the sampling distribution.}
\label{Plot1}
\end{figure}

\begin{figure}[!ht]
\centering
\includegraphics[width=\textwidth]{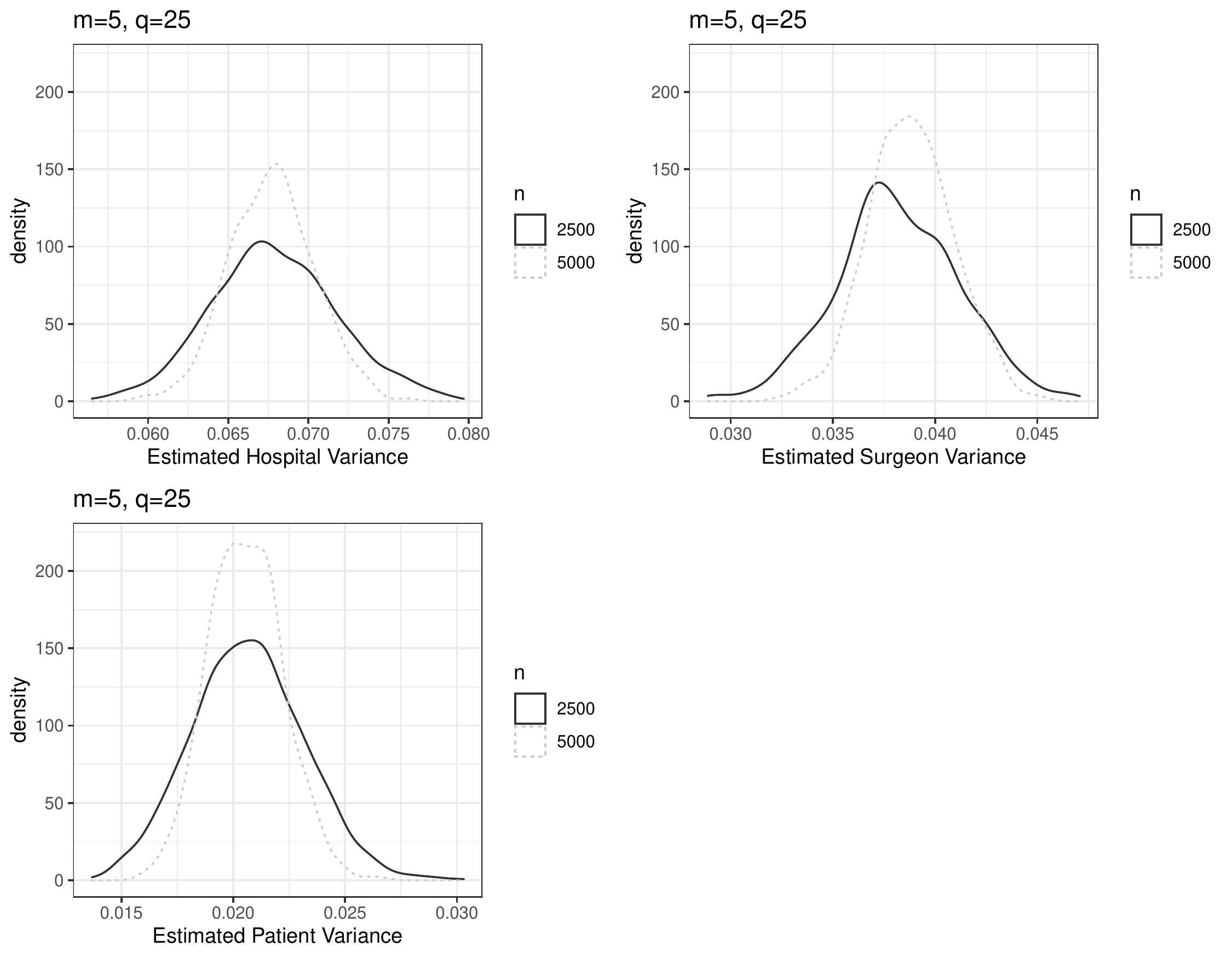}
\caption{Density plots for the simulated sampling distributions of the case-mix, between-hospital, and between-surgeon variance components with a binary outcome, based on 1000 replications.}
\label{Plot2}
\end{figure}

The bars in Figure \ref{Plot1} show the simulated sampling distribution means for the three variance components for the binary outcomes under different combinations of $n$ (total number of patients), $m$ (total number of hospitals), and $q$ (total number of surgeons), based on 1000 replications. The $95\%$ quantile interval for the sampling distribution is represented by the black error bar. The $95\%$ confidence interval for the mean is represented by the blue error bar; this reflects the Monte Carlo error in the estimated mean of the sampling distribution. The red dots indicate the true values of the variance components. From the results, we can observe that the between-hospital and case-mix components are well estimated under all scenarios. Accurate estimation of the within-hospital between-surgeon component requires sufficient surgeon-specific patient volumes, which are the largest under the scenario $n=5000$, $m=5$ and $q=25$, explaining the more precise estimate for this variance component therein. Figure \ref{Plot2} shows density plots for the simulated sampling distributions of the three variance components with varying $n$ and fixed $m$ and $q$. These are fairly normal-shaped, and demonstrate decreasing variability with increasing number of patients.

\begin{figure}[!h]
        \centering
        \subfloat[]{\label{Plot3}\includegraphics[width=0.5\textwidth]{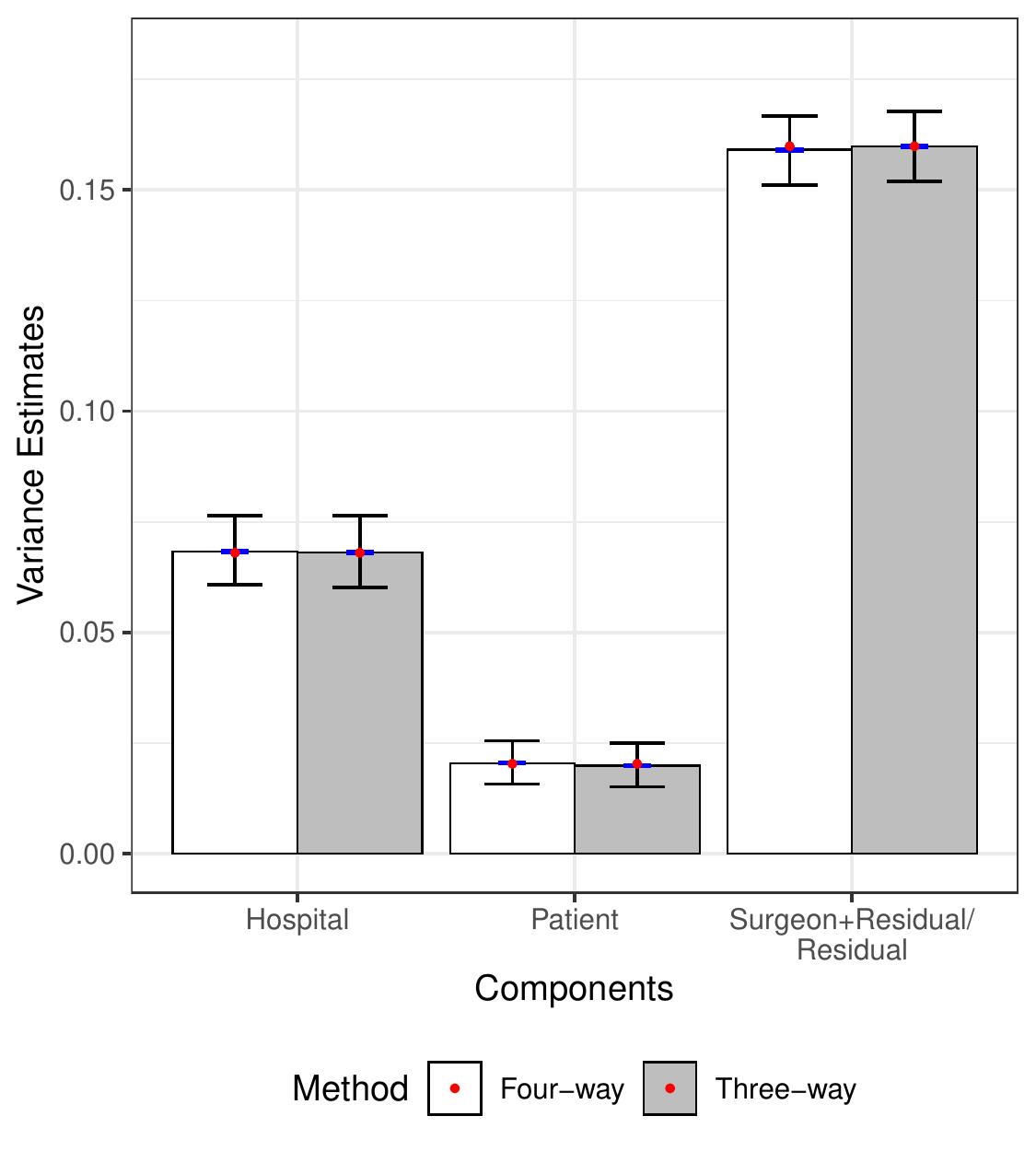}}
        \subfloat[]{\label{Plot4}\includegraphics[width=0.5\textwidth]{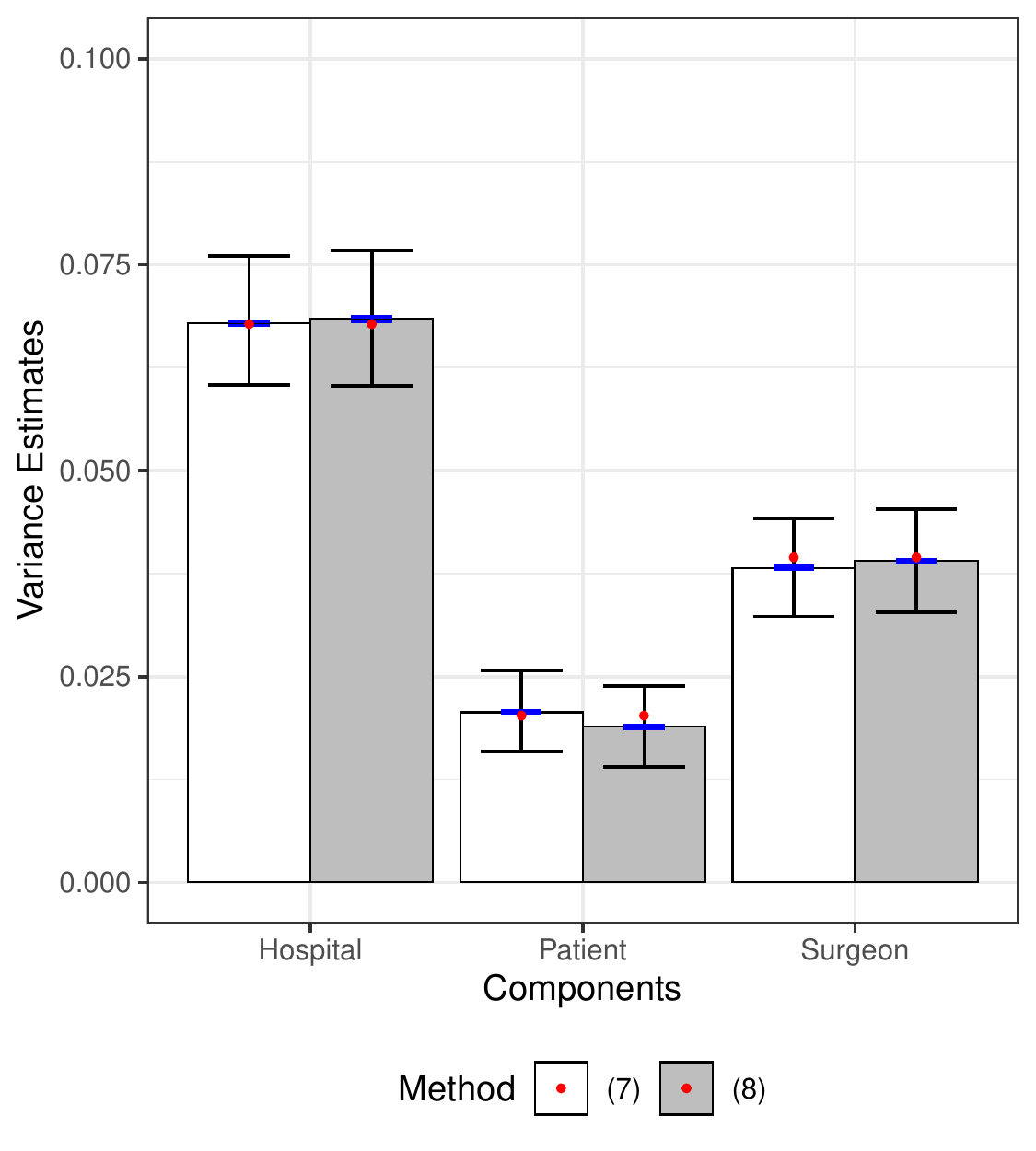}}
        \caption{Panel (a): Simulated sampling distribution means for the variance components of the three-way decomposition \eqref{intro1}, and the same components estimated through the four-way decomposition \eqref{eq7}. The red dots indicate the true variances. The $95\%$ quantile interval of the sampling distribution is represented by the black error bar. The $95\%$ confidence interval for the mean is represented by the blue error bar, reflecting the Monte Carlo error in the estimated mean of the sampling distribution. Panel (b): Simulated sampling distribution means for the three variance components (without residual variance) estimated using the model-based formulation \eqref{eq7} and the alternative semi-parameric formulation \eqref{eq8}, based on 1000 replications. The red dots indicate the true variances. The $95\%$ quantile interval of the sampling distribution is represented by the black error bar. The $95\%$ confidence interval for the mean is represented by the blue error bar, reflecting the Monte Carlo error in the estimated mean of the sampling distribution.}
\end{figure}

The gray bars in Figure \ref{Plot3} show the simulated sampling distribution means for the estimated components of the three-way decomposition \eqref{intro1}, and the white bars show the corresponding components obtained through the four-way decomposition \eqref{eq7} by adding up the third (between surgeon) and fourth (residual) variance components. The estimates are similar, demonstrating that the new between-surgeon variance component is part of the residual variance in the three-way decomposition.

The gray bars in Figure \ref{Plot4} show the simulated sampling distribution means for the three variance components using the model-based formulation \eqref{eq7} and the white bars the alternative semi-parameric formulation \eqref{eq8}. The point estimates, as well as their variability are similar under both approaches, demonstrating that both are appropriate for the point estimation. 

\section{Discussion}\label{section:discussion}

The methods in the present paper are aimed at helping to assess the usefulness of a given process or outcome in constructing a disease-specific quality indicator for identifying performance related between hospital and between surgeon variation. Although here we focused on two levels of hierarchical clustering due to the motivating application, the proposed four-way variance decomposition proposed here could be generalized to arbitrary number of levels of hierarchical clustering, by introducing further conditioning variables. The additional variance components will come out of the residual variation for lower level clusters introduced, as we observed for the surgeons within hospitals. While in the present context there are no more lower level clusters to introduce, we could introduce higher level clusters such as the Local Health Integration Networks (LHINs) which are health administrative subregions in Ontario.

For estimation of the variance decompositions, we used nested random effect models, as these can easily accommodate large number of small categories without identifiability problems that would be present with corresponding fixed effect models. For simplicity, we also omitted hospital-case-mix and surgeon-case-mix interaction terms from the models; however, in principle these can be easily incorporated, either through fixed or random effects, as the interpretation of the variance decompositions is separate from the parametrization of the hospital and surgeon effects. In fact, the formulation of the causal variance decompositions does not dictate what kind of models are used to estimate the predictive means/probabilities needed for calculation of the decomposition. Instead of parametric models, predictions derived through machine learning algorithms might be useful as well, though it is an open question how well these can capture the effects of large number of levels in the categorical exposures. In the context of multi-category categorical exposures, the components in the causal variance decomposition can be seen as a way to concisely summarize a large number of pairwise causal contrasts. In principle the same approach could also be extended to other types of exposures, including continuous and function valued exposures, which is one further research direction we are pursuing.

We borrowed nested potential outcome notation from causal mediation analysis and instrumental variable estimation literature to represent the nested exposure levels. However, although the path hospital $\rightarrow$ surgeon $\rightarrow$ outcome in the causal diagram \ref{figure:dag} resembles mediation, we note that the current problem is not a mediation problem, due to the surgeons being nested within the hospitals by definition in our analysis. Because of this, the surgeon effect is separate from the hospital effect, rather than a component of it. However, in \citet{daignault2019causal} we proposed methodology for mediation analysis in the quality of care context, with the aim of quantifying how much of between hospital differences in an outcome type measure could be accounted for by of between hospital differences in a process type measure considered as a mediator, using the hospital $\rightarrow$ minimally invasive surgery $\rightarrow$ length of stay pathway as an example. This raises the question of decomposing between hospital variation in a mediation analysis sense. This problem has connections to various $R^2$ and effect size type measures that have been suggested in the psychometric literature for measuring mediation in the linear structural equation modeling framework \citep[e.g.][]{de2012r,lachowicz2018novel,miovcevic2018statistical}. We are currently working on extending the causal variance decomposition approach to allow for decomposing between hospital variance into direct and indirect effects.

\subsection*{Acknowledgement}

This work was supported by a Discovery Grant from the Natural Sciences and Engineering Research Council of Canada (to OS) and the Ontario Institute for Cancer Research through funding provided by the Government of Ontario (to BC).

\end{document}